# Electromagnetic radiation and the self torque of an oscillating magnetic dipole

Masud Mansuripur† and Per K. Jakobsen‡

†College of Optical Sciences, The University of Arizona, Tucson, Arizona, USA
‡Department of Mathematics and Statistics, UIT The Arctic University of Norway, Tromsø, Norway

**Abstract**. A uniformly-charged spherical shell of radius $R$, mass $m$, and total electrical charge $q$, having an oscillatory angular velocity $\boldsymbol{\Omega}(t)$ around a fixed axis, is a model for a magnetic dipole that radiates an electromagnetic field into its surrounding free space at a fixed oscillation frequency $\omega$. An exact solution of the Maxwell-Lorentz equations of classical electrodynamics yields the self-torque of radiation resistance acting on the spherical shell as a function of $R$, $q$, and $\omega$. Invoking the Newtonian equation of motion for the shell, we relate its angular velocity $\boldsymbol{\Omega}(t)$ to an externally applied torque, and proceed to examine the response of the magnetic dipole to an impulsive torque applied at a given instant of time, say, $t = 0$. The impulse response of the dipole is found to be causal down to extremely small values of $R$ (i.e., as $R \to 0$) so long as the exact expression of the self-torque is used in the dynamical equation of motion of the spherical shell.

**1. Introduction**. In a recent paper,[1] we examined the electromagnetic (EM) radiation by a small spherical electric dipole and pointed out the significance of the role played by the exact radiation reaction function in the context of the causal behavior of the dipole's response to external excitations. The present paper extends the results of [1] to an oscillating magnetic dipole modelled as a rotating electrically-charged spherical shell. While the general conclusions of the present paper parallel those of the previous one, we believe the magnetic dipole's inherent advantages over an electric dipole make it worthy of its own separate analysis. Whereas an electric dipole's positive and negative charges physically separate from each other during each oscillation period, the magnetic dipole's rotary motion does not entail a similar separation of charges. This feature of the magnetic dipole not only removes restrictions on its oscillation amplitude, but also eliminates the restoring force that the opposite charges of an electric dipole, continually and unavoidably, exert upon each other. Such simplifications and reductions in the number of physical constraints on the system under investigation enable one to focus attention on the salient features of the system that impact its causal or acausal behavior. Another difference between the two dipoles is that the inertial masses of the charged particles constituting an electric dipole remain fixed as the dipole's radius is made to approach zero, whereas in the case of a magnetic dipole, the moment of inertia of the rotating particle diminishes along with its shrinking radius. All in all, an oscillating spherical shell imitating a magnetic dipole provides a simpler model for studying the causal or acausal behavior of small electric charges in the limit when their dimensions are made to approach zero.

The organization of the paper is as follows. In Sec.2, we describe the spinning spherical shell model of a classical magnetic dipole, and derive exact expressions for the single-frequency vector potential in the free space regions inside and outside the sphere. The calculated vector potential is used in Sec.3 to arrive at rigorous expressions for the EM fields surrounding the shell, and also to compute the rate of EM energy radiation as a function of the oscillation frequency $\omega$ for a given radius $R$ and electric charge $q$ of the spherical dipole. The self $E$-field of the dipole is then used in Sec.4, in conjunction with Newton's second law of motion, to relate the angular velocity $\boldsymbol{\Omega}(t) = \Omega_0\hat{\boldsymbol{z}}e^{-\mathrm{i}\omega t}$ of the spherical shell to an externally applied torque $\boldsymbol{T}(t) = T_0\hat{\boldsymbol{z}}e^{-\mathrm{i}\omega t}$ that drives the oscillations of the charged sphere at the desired frequency $\omega$. The end result of this section is an expression for the transfer function $\Omega_0/T_0$ of the dipole as a function of its excitation frequency $\omega$ for arbitrary values of the radius $R$, the overall charge $q$, and the total mass $m$ of the spinning spherical shell.

Section 5 is devoted to a discussion of the role played by the poles of the transfer function $\Omega_0/T_0$ in the causal response of our magnetic dipole to an impulsive excitation. Specifically, we argue that the presence of any number of poles in the upper-half of the complex $\omega$-plane provides a clear indication that the impulse-response of the dipole is acausal. Here, we also show numerical results that confirm that, while the small-radius approximation to the self-torque (i.e., radiation resistance) leads to the prediction of acausal behavior, the exact self-torque places all the poles of $\Omega_0/T_0$ in the lower-half plane, thus ensuring the dipole's causal response.

In Sec.6, we argue that the radiation reaction function $\Gamma(\omega)$ can be split into two parts: (i) a part that is in-phase with the oscillating electric current around the spherical shell and, therefore, accounts for the radiated EM energy; and (ii) a part that is out-of-phase with the electric current and can be associated with the underlying mechanism that drives the internal exchange between a "material component" and an "EM component" of the mass of the dipole. Considering that the dipole's inertial mass $m$ is already taken into account through its contribution to the moment of inertia of the sphere, it is tempting to remove the out-of-phase component of the radiation reaction function from the equation of motion—ostensibly because its effect has already been accounted for through the use of a fixed moment of inertia for the particle. However, it will be shown in Sec.6 that removing even a small fraction of the out-of-phase component of $\Gamma(\omega)$ brings about acausal behavior by putting an infinite number of poles into the upper-half plane of the argument $\omega$ of the transfer function. A brief discussion of this curious behavior of the transfer function $\Omega_0/T_0$ is relegated to the final section of the paper.

**2. Model of magnetic dipole as an oscillating electrical current around a spherical shell**. Figure 1 shows a uniformly-charged spherical shell of radius $R$ and total electrical charge $q$, spinning with a time-dependent angular velocity around the $z$-axis. Let the total mass $m$ of the shell be uniformly distributed over its surface area. The moment of inertia $I$ of the shell is readily computed in the spherical $(r, \theta, \varphi)$ coordinate system, as follows:

$$I = \int_{\theta=0}^{\pi} \left(\frac{m}{4\pi R^2}\right)(R\sin\theta)^2(2\pi R^2 \sin\theta)\mathrm{d}\theta = \tfrac{2}{3}mR^2. \tag{1}$$

Defining $\Omega_0 = |\Omega_0|e^{i\phi_0}$ as the complex amplitude of the sinusoidal oscillations around the $z$-axis, we write the angular velocity of the spherical shell as follows:

$$\dot{\varphi}(t)\hat{\mathbf{z}} = |\Omega_0|\cos(\omega t - \phi_0)\hat{\mathbf{z}} = \mathrm{Re}(\Omega_0 e^{-i\omega t})\hat{\mathbf{z}} = \mathrm{Re}[\boldsymbol{\Omega}(t)]. \tag{2}$$

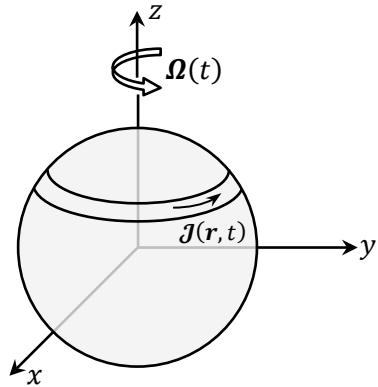

**Fig.1**. A thin spherical shell of radius $R$, mass $m$, and total electric charge $q$ rotates around the $z$-axis with angular velocity $\boldsymbol{\Omega}(t)$. The current density $\boldsymbol{\mathcal{J}}(\mathbf{r}, t)$ is maximum at the equator and drops to zero (in proportion to $\sin\theta$) as the polar angle $\theta$ approaches zero at the north pole, and $\pi$ at the south pole.

Introducing $\mathcal{J}_{s0} = q\Omega_o/(4\pi R)$ as the (complex) amplitude of the $\varphi$-directed surface current density at the equator, the complete expression of the surface-current-density will be

$$\boldsymbol{J}_s(r = R, \theta, \varphi, t) = \left(\frac{q}{4\pi R^2}\right)(R\sin\theta)\Omega(t)\widehat{\boldsymbol{\varphi}} = \mathcal{J}_{s0}\sin\theta\, e^{-i\omega t}\widehat{\boldsymbol{\varphi}} = \left(\frac{q}{4\pi R}\right)\boldsymbol{\Omega}(t)\times\hat{\boldsymbol{r}}. \tag{3}$$

The magnetic dipole moment of the spinning sphere is thus given by[2]

$$\boldsymbol{m}(t) = \mu_o \int_{\theta=0}^{\pi}(\pi R^2\sin^2\theta)\left(\frac{q}{4\pi R^2}\right)(R\sin\theta\,\boldsymbol{\Omega})R\mathrm{d}\theta = \tfrac{1}{3}\mu_o q R^2\boldsymbol{\Omega}(t), \tag{4}$$

where $\mu_o = 4\pi\times 10^{-7}$ henry/meter is the permeability of free space in the SI system of units. Writing $\boldsymbol{m}(t) = m_o e^{-i\omega t}\hat{\boldsymbol{z}}$, the complex dipole moment amplitude is $m_o = \mu_o(4\pi R^3/3)\mathcal{J}_{s0}$. We now write the current density of the spinning sphere as follows:

$$\boldsymbol{J}(\boldsymbol{r},t) = \mathcal{J}_{s0}\delta(r-R)\sin\theta\, e^{-i\omega t}\widehat{\boldsymbol{\varphi}} = \mathcal{J}_{s0}\delta(r-R)e^{-i\omega t}\,\hat{\boldsymbol{z}}\times\hat{\boldsymbol{r}}. \tag{5}$$

Appendix A shows that the vector potential produced inside and outside the shell in accordance with the standard (i.e., Maxwellian) theory of electrodynamics are given by[2-7]

$$\boldsymbol{A}_{\text{in}}(\boldsymbol{r},t) = \frac{\mu_o\mathcal{J}_{s0}\sin\theta\,(1 - iR\omega/c)[\sin(r\omega/c) - (r\omega/c)\cos(r\omega/c)]}{r^2(\omega/c)^3}e^{-i\omega(t-R/c)}\,\widehat{\boldsymbol{\varphi}}. \tag{6}$$

$$\boldsymbol{A}_{\text{out}}(\boldsymbol{r},t) = \frac{\mu_o\mathcal{J}_{s0}\sin\theta\,(1 - ir\omega/c)[\sin(R\omega/c) - (R\omega/c)\cos(R\omega/c)]}{r^2(\omega/c)^3}e^{-i\omega(t-r/c)}\,\widehat{\boldsymbol{\varphi}}. \tag{7}$$

Here, $c = 1/\sqrt{\mu_o\varepsilon_o}$ is the speed of light in vacuum, with $\mu_o$ and $\varepsilon_o$ being the permeability and permittivity of free space. Note the continuity of the vector potential at the sphere's surface, as well as its compliance with the Lorenz gauge, which, in the absence of a scalar potential, requires that $\boldsymbol{\nabla}\cdot\boldsymbol{A}$ be zero.

**3. Electric and magnetic fields**. Having found the vector potential, the $H$-field is derived from the standard relation $\mu_o\boldsymbol{H}(\boldsymbol{r},t) = \boldsymbol{\nabla}\times\boldsymbol{A}(\boldsymbol{r},t)$ and, in the absence of a scalar potential, the $E$-field is obtained from $\boldsymbol{E}(\boldsymbol{r},t) = -\partial_t\boldsymbol{A}(\boldsymbol{r},t)$.[2,3] Denoting the impedance of free space by $Z_o$, where $Z_o = \sqrt{\mu_o/\varepsilon_o}$, we will have

$$\boldsymbol{E}_{\text{in}}(\boldsymbol{r},t) = \frac{Z_o\mathcal{J}_{s0}\sin\theta\,[(R\omega/c) + i][\sin(r\omega/c) - (r\omega/c)\cos(r\omega/c)]}{(r\omega/c)^2}e^{-i\omega(t-R/c)}\widehat{\boldsymbol{\varphi}}. \tag{8}$$

$$\boldsymbol{E}_{\text{out}}(\boldsymbol{r},t) = \frac{Z_o\mathcal{J}_{s0}\sin\theta\,[(r\omega/c) + i][\sin(R\omega/c) - (R\omega/c)\cos(R\omega/c)]}{(r\omega/c)^2}e^{-i\omega(t-r/c)}\widehat{\boldsymbol{\varphi}}. \tag{9}$$

$$\boldsymbol{H}_{\text{in}}(\boldsymbol{r},t) = \mathcal{J}_{s0}(1 - iR\omega/c)$$
$$\times\left\{\frac{[\sin(r\omega/c) - (r\omega/c)\cos(r\omega/c)](2\cos\theta\,\hat{\boldsymbol{r}} + \sin\theta\,\hat{\boldsymbol{\theta}})}{(r\omega/c)^3} - \frac{\sin(r\omega/c)\sin\theta\,\hat{\boldsymbol{\theta}}}{r\omega/c}\right\}e^{-i\omega(t-R/c)}. \tag{10}$$

$$\boldsymbol{H}_{\text{out}}(\boldsymbol{r},t) = \mathcal{J}_{s0}[\sin(R\omega/c) - (R\omega/c)\cos(R\omega/c)]$$
$$\times\left[\frac{(1 - ir\omega/c)(2\cos\theta\,\hat{\boldsymbol{r}} + \sin\theta\,\hat{\boldsymbol{\theta}})}{(r\omega/c)^3} - \frac{\sin\theta\,\hat{\boldsymbol{\theta}}}{r\omega/c}\right]e^{-i\omega(t-r/c)}. \tag{11}$$

Note that, while $E_\varphi$ and $H_r$ are continuous at the shell surface, the discontinuity of the tangential $H$ at $r = R$ is precisely matched by the surface current density. The time-averaged rate of energy flow outside the sphere is readily found to be

$$\langle \boldsymbol{S}_{\text{out}}(\boldsymbol{r},t) \rangle = \tfrac{1}{2}\text{Re}(\boldsymbol{E}_{\text{out}} \times \boldsymbol{H}_{\text{out}}^*) = \frac{Z_0|\mathcal{J}_{s0}|^2[\sin(R\omega/c) - (R\omega/c)\cos(R\omega/c)]^2 \sin^2\theta \hat{\boldsymbol{r}}}{2(r\omega/c)^2}. \quad (12)$$

Integrating Eq.(12) over a spherical surface of arbitrary radius $r$ yields

$$\text{Emitted power} = \int_{\theta=0}^{\pi} 2\pi r^2 \sin\theta \, \langle \boldsymbol{S}_{\text{out}}(\boldsymbol{r},t) \rangle \, \mathrm{d}\theta = \frac{4\pi Z_0|\mathcal{J}_{s0}|^2 [\sin(R\omega/c) - (R\omega/c)\cos(R\omega/c)]^2}{3(\omega/c)^2}. \quad (13)$$

It is easy to show that the emitted power equals the negative of the time-averaged work done by the self-field on the surface current of the shell;[6,7] that is,

$$\int_{\theta=0}^{\pi} \tfrac{1}{2}\text{Re}\big[\boldsymbol{E}_{\text{self}}(r=R,\theta,\varphi,t) \cdot \mathcal{J}_{s0}^* \sin\theta \, e^{i\omega t} \hat{\boldsymbol{\varphi}}\big] 2\pi R^2 \sin\theta \, \mathrm{d}\theta$$

$$= -\frac{4\pi Z_0|\mathcal{J}_{s0}|^2 [\sin(R\omega/c) - (R\omega/c)\cos(R\omega/c)]^2}{3(\omega/c)^2}. \quad (14)$$

For sufficiently small values of $R\omega/c$, the Taylor series expansions $\sin x = x - x^3/3! + \cdots$ and $\cos x = 1 - x^2/2! + \cdots$ lead to the approximate expression $|m_0|^2\omega^4/(12\pi Z_0 c^2)$ for the time-averaged EM power emitted by the dipole moment of amplitude $m_0 = \mu_0(4\pi R^3/3)\mathcal{J}_{s0}$ oscillating at the frequency $\omega$.

**4. Response of the spherical magnetic dipole to an externally applied torque**. Let the external torque $\boldsymbol{T}(t) = T_0 e^{-i\omega t}\hat{\boldsymbol{z}}$ act on our magnetic dipole.[†] The dipole responds by acquiring an angular velocity $\boldsymbol{\Omega}(t)$, which oscillates with the frequency $\omega$ of the applied torque. Using Eq.(8) or Eq.(9), we compute the self torque of radiation resistance (produced by the radiated $E$-field) acting on the uniformly-charged spherical shell, as follows:

$$\boldsymbol{T}_{\text{self}}(t) = \hat{\boldsymbol{z}}\int_{\theta=0}^{\pi} \left(\frac{q}{4\pi R^2}\right) E_{\text{self}}(r=R,\theta,\varphi,t)(R\sin\theta)(2\pi R^2 \sin\theta)\mathrm{d}\theta$$

$$= \left(\frac{2qR}{3}\right)\frac{Z_0 \mathcal{J}_{s0} \, [(R\omega/c) + i] \, [\sin(R\omega/c) - (R\omega/c)\cos(R\omega/c)]}{(R\omega/c)^2} e^{-i\omega(t-R/c)}\hat{\boldsymbol{z}}$$

$$= \left(\frac{Z_0 q^2 \Omega_0}{6\pi}\right)\frac{[\sin(R\omega/c) - (R\omega/c)\cos(R\omega/c)] \times [(R\omega/c)+i]e^{iR\omega/c}}{(R\omega/c)^2} e^{-i\omega t} \hat{\boldsymbol{z}}. \quad (15)$$

Newton's second law may now be invoked to write the dynamic equation of motion for the shell. In the presence of a dynamic friction torque (friction coefficient $= \beta$), we will have

$$\boldsymbol{T}(t) + \boldsymbol{T}_{\text{self}}(t) - \beta\boldsymbol{\Omega}(t) = I\dot{\boldsymbol{\Omega}}(t). \quad (16)$$

---

[†]Suppose the driving agent is a spatially uniform magnetic field $\boldsymbol{H}(t) = H_0\hat{\boldsymbol{z}}\cos(\omega t)$. For a ring of the shell at the polar coordinate $\theta$, Maxwell's equation $\boldsymbol{\nabla} \times \boldsymbol{E} = -\partial_t \boldsymbol{B}$ yields $2\pi R \sin\theta \, E_\varphi = \pi(R\sin\theta)^2\mu_0 H_0 \omega \sin(\omega t)$, or $E_\varphi(\theta,t) = \tfrac{1}{2}\mu_0 R H_0 \omega \sin\theta \sin(\omega t)$. The torque acting on the uniformly charged spherical shell will then be

$$T_z(t) = \int_{\theta=0}^{\pi}(q/4\pi R^2)E_\varphi(\theta,t)(R\sin\theta)(2\pi R^2 \sin\theta)\mathrm{d}\theta = \tfrac{1}{3}\mu_0 q R^2 H_0 \omega \sin(\omega t).$$

This, of course, is an approximation, as the other relevant Maxwell equation, $\boldsymbol{\nabla} \times \boldsymbol{H} = \varepsilon_0 \partial_t \boldsymbol{E}$, has *not* been considered in this derivation. Nevertheless, it demonstrates the feasibility of generating a contactless EM torque.

Substitution from the preceding equations into Eq.(16) yields the following transfer function for the system:

$$\frac{\Omega_0}{T_0} = \frac{i}{I\omega + \Gamma(\omega) + i\beta}, \quad (17)$$

where

$$\Gamma(\omega) = \left(\frac{Z_0 q^2}{6\pi}\right) \frac{[\sin(R\omega/c) - (R\omega/c)\cos(R\omega/c)] \times (1 - iR\omega/c)e^{iR\omega/c}}{(R\omega/c)^2}. \quad (18)$$

For sufficiently small values of $R\omega/c$, the Taylor series expansions $\sin x = x - x^3/3! + \cdots$ and $\cos x = 1 - x^2/2! + \cdots$ lead to the following approximate expression for the radiation reaction function:

$$\Gamma(\omega) \cong \left(\frac{Z_0 q^2}{18\pi}\right)[(R\omega/c) + \tfrac{2}{5}(R\omega/c)^3 + \tfrac{1}{3}i(R\omega/c)^4]. \quad (19)$$

In this approximation, it is clear that radiation reaction contributes additive terms to both the moment of inertia $I$, and the friction coefficient $\beta$ as they appear in Eq.(17).[‡] Depending on the various parameter values, it is conceivable for the transfer function of Eq.(17) with the approximate $\Gamma(\omega)$ of Eq.(19) to have one or more poles in the upper-half of the $\omega$-plane, thereby rendering the system acausal. This, in fact, turns out to be the case for the broad range of parameter values examined in the next section. An important question for the numerical analysis taken up in Sec.5 is whether Eq.(17), in conjunction with the *exact* radiation reaction function of Eq.(18), could exhibit causal behavior.

In the following analysis the total charge and total mass of the spherical shell will be assumed to be those of a free electron, namely $q = -1.6 \times 10^{-19}$ C and $m = 9.11 \times 10^{-31}$ kg. Our equation of motion will not change if we imagine a system consisting of two oppositely charged spherical shells (one immediately inside and essentially in contact with the other) that have equal and opposite charges $\pm q/2$, equal masses $m/2$, equal friction coefficients $\beta/2$, and equal but opposite angular velocities $\pm\boldsymbol{\Omega}(t)$. The particle now becomes charge-neutral but, since its opposite (internal) charges rotate in opposite directions, it will have the same overall magnetic dipole moment $\boldsymbol{m}(t) = \tfrac{1}{3}\mu_0 q R^2 \boldsymbol{\Omega}(t)$ as before. Thus, with a judicious choice of the parameter values $q$, $m$, and $R$, Eqs.(17)-(19) can be applied not only to a charged particle, but also to a neutral particle such as neutron ($m_{\text{neutron}} = 1.675 \times 10^{-27}$ kg, $R_{\text{neutron}} \cong 0.8$ fm).

**5. Causality and the absence of poles in the upper-half plane**. The transfer function of Eq.(17) is the Fourier transform of the impulse-response of the spherical magnetic dipole presently under consideration. As we have argued in [1], if one or more poles of this transfer function happen to be in the upper-half of the complex $\omega$-plane, the impulse-response will have nonzero values *before* the arrival (in time) of the externally-applied torque $\boldsymbol{T}(t) = T_0 \delta(t)\hat{\boldsymbol{z}}$. This, of course, is a clear indication that the response of the dipole to (externally applied) driving torques is acausal.

---

[‡] The first term of the approximate $\Gamma(\omega)$ in Eq.(19) increments the mechanical moment of inertia $I = \tfrac{2}{3}mR^2$ by an EM contribution equal to $\mu_0 q^2 R/(18\pi)$. Thus, the non-EM contribution to the overall mass $m$ of the particle must be negative if $R$ shrinks below $\mu_0 q^2/(12\pi m)$. This critical radius is of the same order of magnitude as the classical radius of a charged particle [1,2]. (A similar argument also applies to a charge-neutral particle consisting of two identical spherical shells of equal and opposite charge rotating in opposite directions, with one shell immediately inside the other.) It is customary to resort to a mass-renormalization scheme by reducing the mass $m$ of the particle in order to compensate for the electrodynamic contribution to the inertial mass.[8] In the context of the present paper, mass-renormalization would entail subtracting $\mu_0 q^2/(12\pi R)$ from the mass $m$. However, since we are not convinced that this is the best way to handle the electrodynamic contribution to the inertial mass, we eschew this approach to mass-renormalization in favor of the alternative scheme that is discussed in Sec.6.

In the complex $\omega$-plane depicted in Fig.2(a), red dots mark the locations of the four poles of the transfer function of Eq.(17), when the small-radius approximation to $\Gamma(\omega)$ of Eq.(19) is used along with the parameter values $q = -1.6 \times 10^{-19}$ C, $m = 9.11 \times 10^{-31}$ kg (corresponding to a single electron), $R = 1.0$ nm, and $\beta = 0$. Two of the poles are seen to be in the upper-half of the $\omega$-plane, thus revealing the acausal nature of the impulse-response. Shrinking the radius $R$ down to 1.0 fm, or increasing the mass $m$ (e.g., using $m_{neutron}$ in place of $m_{electron}$), or raising $\beta$ by as much as $10^{-10}$ kg·m²/s did not make the upper poles move into the lower half plane.

In contrast, Fig.2(b) shows that, for the exact radiation reaction function $\Gamma(\omega)$ of Eq.(18), the transfer function $\Omega_0/T_0$ of Eq.(17) possesses an infinite number of poles, all residing in the lower-half of the $\omega$-plane. The exact location of the poles, of course, varies with the system parameters $(R, q, m, \beta)$, but our numerical studies indicate that the impulse-response in this case remains causal over a broad range of the parameter values. (As was done in Ref.[1], we also used Cauchy's argument principle to confirm that the poles of the transfer function remain in the lower-half of the $\omega$-plane.) We mention in passing that, in the case of $\beta = 0$, the half-residue of the first-order pole at $\omega = 0$ adds a constant term to the impulse-response during the time-interval $t < 0$. This, however, is not indicative of acausal behavior, but rather a reminder that the initial condition of the dipole can be adjusted to ensure that $\Omega(t) = 0$ for $t < 0$.

The conclusion is that the predicted acausal behavior based on the approximate $\Gamma(\omega)$ of Eq.(19) is not a reliable indicator of the actual response of our magnetic dipole to an impulsive excitation. When the exact form of the radiation reaction function given by Eq.(18) is used in the calculations, the dipole is found to respond in a causal way.

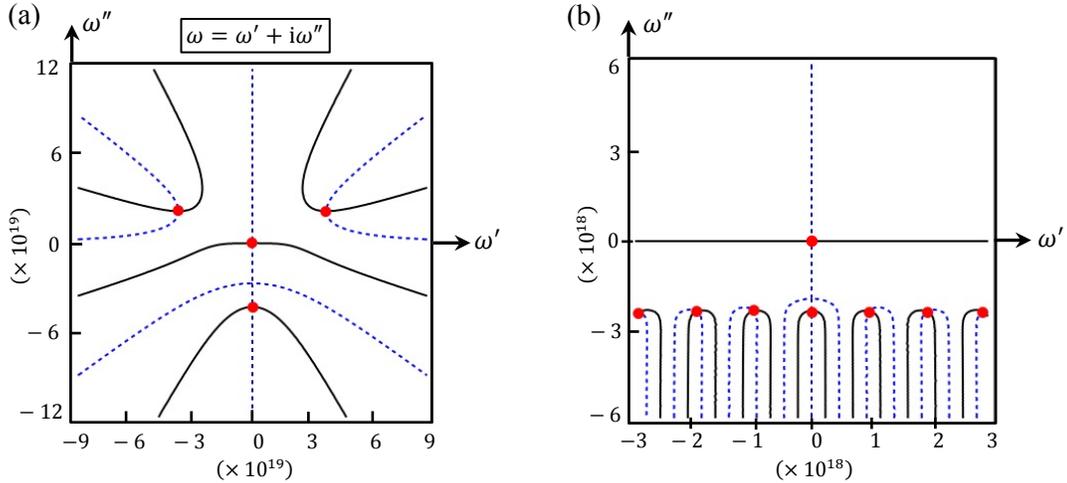

**Fig.2**. Complex-plane diagrams showing the zero contours of the real part (dashed blue) and imaginary part (solid black) of the denominator of Eq.(17); the marked crossing points are the poles of the transfer function $\Omega_0/T_0$. (a) In the case of the approximate $\Gamma(\omega)$ of Eq.(19), aside from the trivial pole at $\omega = 0$, there reside one purely imaginary pole in the lower half plane, and two (symmetrically-positioned with respect to the imaginary axis) poles in the upper half plane. Locations of the nonzero poles vary with $R$, but as $R \to 0$, there always remain two symmetrically-positioned poles in the upper-half plane and one imaginary pole in the lower-half plane. When $m_{neutron}$ is substituted for $m_{electron}$, the general pattern of the pole locations remains the same, although the nonzero poles move further apart. (b) In the case of the exact $\Gamma(\omega)$ of Eq.(18), an infinite number of poles are symmetrically-distributed (again, with respect to the imaginary axis) in the lower half plane. Aside from the trivial pole at $\omega = 0$, none of these poles coincide with those of the approximate transfer function depicted in (a). Locations of the nonzero poles vary with $R$, but as $R \to 0$, the poles remain in the lower-half plane and retain their symmetry with respect to the imaginary axis.

The standard way to infer the causality of the impulse-response from the absence of poles in the upper-half of the $\omega$-plane is to begin by noting that $\Omega_0/T_0 \to 0$ when $\omega \to \infty$ in the upper-half plane. For $t < 0$, the inverse Fourier transform integral of $(\Omega_0/T_0)e^{-i\omega' t}$ over the real axis $\omega'$ is then equated with the integral over a large upper-half semi-circle plus the sum of the residues at the upper-half poles. Considering that the integral over the (infinitely large) semi-circle vanishes, the absence of poles in the upper-half plane heralds the vanishing of the impulse-response over the interval $t < 0$.

A less formal, but perhaps more intuitive, way to arrive at the same conclusion is to begin by supposing that the dipole's impulse-response is, in fact, causal. If the transfer function happens to have a first-order pole at the origin, (e.g., when the friction coefficient $\beta$ in Eq.(17) is set to zero), one should eliminate this pole by multiplying $\Omega_0/T_0$ with $-i\omega$, which is tantamount to replacing the impulse-response with its own time-derivative, thus avoiding situations in which the impulse-response may have a constant nonzero value during the time interval $t < 0$. Stated differently, our starting assumption here is that the impulse-response (or its time-derivative) is precisely zero for $t < 0$, and is sufficiently well-behaved during $t \geq 0$ to have the Fourier transform function $\Omega_0/T_0$ (or $-i\omega\Omega_0/T_0$ if $\omega = 0$ happens to be a pole).

Now, if we multiply this well-behaved impulse-response (or its time-derivative) by $\exp(-\alpha t)$, where $\alpha$ is some positive real number, the Fourier transform of the product function must also be well-behaved. However, the Fourier transform of the product is just our transfer function $\Omega_0/T_0$ (or $-i\omega\Omega_0/T_0$ if $\omega = 0$ happens to be a pole) evaluated at $\omega' + i\alpha$, that is, on a straight line parallel to the real axis $\omega'$ in the upper-half $\omega$-plane. Since our starting assumption was that the product function is well-behaved and that the positive number $\alpha$ is arbitrary, the transfer function in the upper-half-plane cannot go to infinity. The conclusion is that the presence of even one pole in the upper-half-plane is proof that the impulse-response is acausal.

**6. Bypassing the need for mass-renormalization**? One could argue that the relevant self-torque is due only to that part of the self $E$-field that is in-phase with the surface current; that is,

$$\widetilde{\boldsymbol{E}}_{\text{self}}(r = R, \theta, \varphi, t) = -\frac{Z_0 J_{s0} \sin\theta \, [\sin(R\omega/c) - (R\omega/c)\cos(R\omega/c)]^2}{(R\omega/c)^2} e^{-i\omega t} \widehat{\boldsymbol{\varphi}}. \tag{20}$$

Consequently,

$$\widetilde{\boldsymbol{T}}_{\text{self}}(t) = -\left(\frac{Z_0 q^2 \Omega_0}{6\pi}\right) \frac{[\sin(R\omega/c) - (R\omega/c)\cos(R\omega/c)]^2}{(R\omega/c)^2} e^{-i\omega t} \widehat{\boldsymbol{z}}. \tag{21}$$

This means that $\Gamma(\omega)$ of Eq.(18) should be replaced by the following radiation reaction function:

$$\widetilde{\Gamma}(\omega) = i\left(\frac{Z_0 q^2}{6\pi}\right) \frac{[\sin(R\omega/c) - (R\omega/c)\cos(R\omega/c)]^2}{(R\omega/c)^2}. \tag{22}$$

In this way, $\widetilde{\Gamma}(\omega)$ essentially acts as a friction coefficient — albeit one that, unlike $\beta$ in Eq.(17), is frequency-dependent. In contrast to $\Gamma(\omega)$ of Eq.(18), the fact that $\widetilde{\Gamma}(\omega)$ of Eq.(22) does not have a real part clearly indicates that it does *not* make an undesirable EM contribution to the moment of inertia $I$ in Eq.(17). This is the sense in which mass-renormalization is avoided.

The component of the self-torque of Eq.(15) that constitutes the effective self-torque of Eq.(21) is the part that accounts for the rate of outgoing radiation (i.e., the EM energy that leaves the dipole and propagates away, never to return). We contend that this is a reasonable way to account for the radiation resistance torque. One might inquire as to the role of the remaining part of the total self-torque of Eq.(15) — the part that is 90° out-of-phase with the surface current and, therefore, makes no contribution to the radiation. The answer is that, during each oscillation period (frequency = $\omega$), some EM energy goes out into the surrounding **E** and **B** fields, but subsequently returns to the dipole, so that the net energy going out (or coming in) during a full period is precisely zero. In a nutshell, the out-of-phase component that is being discounted here is that part of the self-torque whose role is to release some EM energy into the surrounding space during one half of each oscillation cycle, then reclaim that energy during the remaining half of the cycle.

Classical electrodynamics contends that a fraction of the inertial mass of a charged particle resides in its surrounding EM field, while the remaining part is in some mysterious "material" stuff, sometimes associated with the Poincaré stresses.[2,3,6] Our conjecture here is that the non-radiated EM energy that goes in and out of the dipole during each cycle is intimately tied to its inertial mass. In other words, if the spinning ball of charge is inclined to convert some of its inertial mass to EM energy during one half of each cycle, then bring that energy back in the form of the "mysterious material stuff" during the remaining half of the cycle, then this is just an internal exchange process between one form of mass and another. Therefore, as far as the overall dynamics is concerned, this internal exchange process is irrelevant and may be ignored. Stated differently, since we are already using a fixed value $m$ for the inertial mass, we should not allow the (internal) mass-conversion-related self-torque to enter into the overall dynamics of the particle through the backdoor. The mass-renormalization procedure described in the literature is intended to cancel this double-counting of the self-field contribution to the inertial mass.[8] Thus, by substituting the effective self-torque of Eq.(21) for the total self-torque of Eq.(15), we endeavor to eliminate the need for the conventional mass-renormalization scheme.

Figure 3(a) shows the computed $\omega$-plane locations of the poles of the transfer function $\Omega_0/T_0$ of Eq.(17) in conjunction with the effective radiation reaction function $\tilde{\Gamma}(\omega)$ of Eq.(22) and the same set of parameters $R, q, m, \beta$ as used in Fig.2(b). An infinite number of poles now appear in the upper-half $\omega$-plane, thus making the impulse-response of the dipole acausal. What is more disheartening is that removing *any* fraction of the out-of-phase component of the self-torque, no matter how small, will have a similar deleterious effect on the dipole's transfer function. The pole-location plot in Fig.3(b) shows that attenuating the out-of-phase component of $\Gamma(\omega)$ by as little as one part in $10^{10}$ causes the transfer function to exhibit an infinite number of poles in the upper-half $\omega$-plane. (Appendix B provides a more detailed discussion of the behavior of the poles in the upper-half $\omega$-plane.) Causality of the impulse-response is thus seen to be a delicate matter that requires the presence of the radiation reaction function $\Gamma(\omega)$ in its entirety as given by Eq.(18). This is not to say that efforts at accounting for the EM contribution to the inertial mass $m$ of the particle should be abandoned, but rather that the role of $m$ in the equation of motion of a charged particle is far more nuanced than might appear at a first glance.

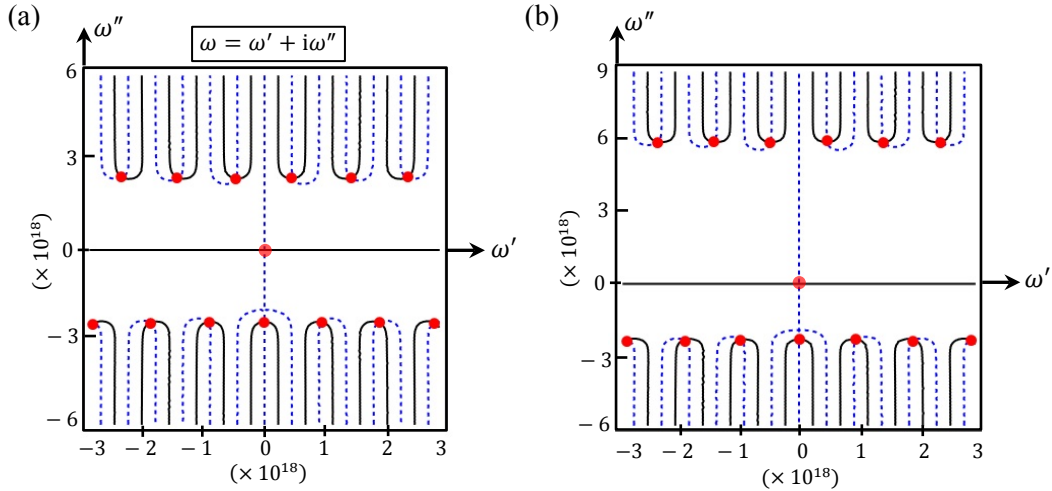

**Fig.3**. (a) Complex-plane diagram showing the zero contours of the real part (dashed blue) and imaginary part (solid black) of the denominator of Eq.(17) with the approximate $\tilde{\Gamma}(\omega)$ of Eq.(22). The marked crossing points are the poles of the transfer function $\Omega_0/T_0$. Here, $q = -1.6 \times 10^{-19}$ C, $m = 9.11 \times 10^{-31}$ kg, $R = 1.0$ nm, and $\beta = 0$. An infinite number of poles, symmetrically-positioned with respect to the imaginary axis, appear in both the upper and lower halves of the $\omega$-plane. The exact locations of the poles vary with $R$, but as $R \to 0$, they retain their symmetry with respect to the imaginary axis and remain in both the upper and lower-halves of the $\omega$-plane. The existence of upper-half-plane poles is evidence that the dipole's response to an impulsive excitation is acausal. (b) Similar to (a) except that the exact $\Gamma(\omega)$ of Eq.(18) is only slightly modified here to attenuate its out-of-phase component by one part in $10^{10}$. The upper-half-plane poles immediately show up even when a tiny fraction of the out-of-phase component of $\Gamma(\omega)$ is taken out.

**7. Concluding remarks**. We have examined the rotary motion of a small, uniformly-charged spherical shell (the classical model of a magnetic dipole), and shown that its predicted response to an externally applied torque is causal provided that the exact form of the radiation reaction function is used in its dynamical equation of motion. When small-radius approximations were used to evince the behavior of the dipole in the limit when it approaches a point-particle, the electro-mechanical response of the particle was found to be acausal. The acausal behavior is thus seen to be a consequence of the approximations used to evaluate the radiation reaction function, rather than heralding a failure of the classical (Maxwell-Lorentz) equations of electrodynamics. These findings are fully accordant with the predicted behavior of the electric dipole that was the subject of our recent paper.[1]

In an attempt to discount the EM contributions to the inertial mass of the particle, we removed a part of the radiation reaction torque that is not directly involved in the extraction of the radiated EM energy from the dipole. The reduced form of the equation of motion, however, immediately sends the predicted response of the particle to external excitations into acausal territory. This is yet another indication that a better understanding is needed of the role of a charged particle's inertial mass in its dynamic equations of motion. Nevertheless, we also remain cognizant of the other shortcomings of the models used in our work in that (i) the analysis has relied on the Newtonian equation of motion, not its relativistic counterpart, and (ii) we have totally ignored quantum mechanics and, in particular, the uncertainty principle that forbids the simultaneous knowledge of the position and momentum of the particle under consideration.

**Acknowledgement**. The authors express their gratitude to Vladimir Hnizdo for generously sharing with us his extensive knowledge of the electrodynamics of charged particles. This work has been supported in part by the AFOSR grant FA9550-19-1-0032.

# Appendix A
## Computing the vector potential of the magnetic dipole

To compute the vector potential $\boldsymbol{A}(\boldsymbol{r},t)$ produced by the electric current distribution $\boldsymbol{J}(\boldsymbol{r},t)$ over the surface of the spherical shell depicted in Fig.1, we begin by Fourier transforming the spatial part of $\boldsymbol{J}(\boldsymbol{r},t)$ given in Eq.(5), as follows:

$$\boldsymbol{J}(\boldsymbol{k}) = \hat{\boldsymbol{z}} \times \int_{r=0}^{\infty}\int_{\vartheta=0}^{\pi} J_{s0}\delta(r-R)\cos\vartheta\,\hat{\boldsymbol{k}}\exp(-\mathrm{i}kr\cos\vartheta)\,2\pi r^2 \sin\vartheta\,\mathrm{d}r\mathrm{d}\vartheta$$

$$= -\frac{\mathrm{i}4\pi J_{s0}[\sin(kR)-kR\cos(kR)]}{k^2}\hat{\boldsymbol{z}}\times\hat{\boldsymbol{k}}. \quad \leftarrow \boxed{\text{Gradshteyn \& Ryzhik}^9 \text{ 3.715}\text{-}11} \tag{A1}$$

Consequently, the contribution of the surface current to the spatial part of the vector potential is

$$\boldsymbol{A}(\boldsymbol{r}) = (2\pi)^{-3}\iiint_{-\infty}^{\infty}\frac{\mu_0 \boldsymbol{J}(\boldsymbol{k})}{k^2-(\omega/c)^2}\exp(\mathrm{i}\boldsymbol{k}\cdot\boldsymbol{r})\,\mathrm{d}\boldsymbol{k}$$

$$= -\frac{\mathrm{i}\mu_0 J_{s0}}{2\pi^2}\hat{\boldsymbol{z}}\times\int_{k=0}^{\infty}\int_{\vartheta=0}^{\pi}\frac{\sin(kR)-kR\cos(kR)}{k^2[k^2-(\omega/c)^2]}\cos\vartheta\,\hat{\boldsymbol{r}}\exp(\mathrm{i}kr\cos\vartheta)\,2\pi k^2\sin\vartheta\,\mathrm{d}k\mathrm{d}\vartheta$$

$$= \frac{2\mu_0 J_{s0}\sin\theta\,\hat{\boldsymbol{\varphi}}}{\pi r^2}\int_0^\infty \frac{[\sin(kR)-kR\cos(kR)]\times[\sin(kr)-kr\cos(kr)]}{k^2[k^2-(\omega/c)^2]}\mathrm{d}k \leftarrow \boxed{\text{Gradshteyn \& Ryzhik}^9 \text{ 3.715}\text{-}11}$$

$$\boxed{\text{residue theorem}}\downarrow$$

$$= \mu_0 J_{s0}\sin\theta\,\hat{\boldsymbol{\varphi}}\begin{cases}\dfrac{[\cos(R\omega/c)+(R\omega/c)\sin(R\omega/c)]\times[\sin(r\omega/c)-(r\omega/c)\cos(r\omega/c)]}{r^2(\omega/c)^3}; & r\leq R,\\[1ex]\dfrac{[\sin(R\omega/c)-(R\omega/c)\cos(R\omega/c)]\times[\cos(r\omega/c)+(r\omega/c)\sin(r\omega/c)]}{r^2(\omega/c)^3}; & r\geq R.\end{cases} \tag{A2}$$

To this, we must now add a contribution from the vacuum field to ensure that the overall vector potential outside the sphere acquires the proper (i.e., retarded) spacetime dependence, namely, $e^{-\mathrm{i}\omega(t-r/c)}$. Introducing the vacuum wavenumber $k_o = \omega/c$ and the as-yet-unspecified vacuum field amplitude $A_o$, we suggest the following spectral distribution for the vacuum field:

$$\boldsymbol{A}_{\text{vac}}(\boldsymbol{k},t) = A_o\delta(k-k_o)e^{-\mathrm{i}\omega t}(\hat{\boldsymbol{z}}\times\hat{\boldsymbol{k}}). \tag{A3}$$

The space part of the vacuum potential will thus be

$$\boldsymbol{A}_{\text{vac}}(\boldsymbol{r}) = (2\pi)^{-3}\int_{k=0}^{\infty}\int_{\vartheta=0}^{\pi}A_o\delta(k-k_o)(\hat{\boldsymbol{z}}\times\hat{\boldsymbol{k}})\exp(\mathrm{i}kr\cos\vartheta)\,2\pi k^2\sin\vartheta\,\mathrm{d}k\mathrm{d}\vartheta$$

$$= (2\pi)^{-2}A_o k_o^2(\hat{\boldsymbol{z}}\times\hat{\boldsymbol{r}})\int_{\vartheta=0}^{\pi}\sin\vartheta\cos\vartheta\exp(\mathrm{i}k_o r\cos\vartheta)\,\mathrm{d}\vartheta$$

$$= \frac{\mathrm{i}A_o}{2\pi^2 r^2}[\sin(r\omega/c)-(r\omega/c)\cos(r\omega/c)]\sin\theta\,\hat{\boldsymbol{\varphi}}. \tag{A4}$$

Comparison with Eq.(A2) for the field outside the sphere ($r\geq R$) now allows us to fix the (heretofore unknown) coefficient $A_o$. We will have

$$\boldsymbol{A}_{\text{vac}}(\boldsymbol{r}) = \frac{\mathrm{i}\mu_0 J_{s0}[\sin(R\omega/c)-(R\omega/c)\cos(R\omega/c)]\times[\sin(r\omega/c)-(r\omega/c)\cos(r\omega/c)]\sin\theta\,\hat{\boldsymbol{\varphi}}}{r^2(\omega/c)^3}. \tag{A5}$$

Combining Eqs.(A2) and (A5), we finally arrive at the total vector potential inside as well as outside the spherical shell, as follows:

$$\boldsymbol{A}_{\text{in}}(\boldsymbol{r},t) = \frac{\mu_0 J_{s0} \sin\theta\ (1 - iR\omega/c)\ [\sin(r\omega/c) - (r\omega/c)\cos(r\omega/c)]}{r^2(\omega/c)^3} e^{-i\omega(t-R/c)}\ \widehat{\boldsymbol{\varphi}}. \tag{A6}$$

$$\boldsymbol{A}_{\text{out}}(\boldsymbol{r},t) = \frac{\mu_0 J_{s0} \sin\theta\ (1 - ir\omega/c)\ [\sin(R\omega/c) - (R\omega/c)\cos(R\omega/c)]}{r^2(\omega/c)^3} e^{-i\omega(t-r/c)}\ \widehat{\boldsymbol{\varphi}}. \tag{A7}$$

As expected, the field outside the sphere as given by Eq.(A7) has the retarded spatio-temporal profile.

## Appendix B
## Approximate formula for the upper-half-plane poles using the Lambert function $W_k(z)$

If we attenuate the *out-of-phase* contribution to $\Gamma(\omega)$ of Eq.(18) by a factor of $(1-\varepsilon)$ while keeping the *in-phase* contribution intact, we will have

$$\Gamma(\omega) \cong \left(\frac{Z_0 q^2}{6\pi}\right) \frac{[\sin(R\omega/c) - (R\omega/c)\cos(R\omega/c)]}{(R\omega/c)^2}$$

$$\times \{(1-\varepsilon)[\cos(R\omega/c) + (R\omega/c)\sin(R\omega/c)] + i[\sin(R\omega/c) - (R\omega/c)\cos(R\omega/c)]\}$$

$$= \frac{Z_0 q^2}{6\pi(R\omega/c)^2} \{(1-\varepsilon)[\tfrac{1}{2}\sin(2R\omega/c) - (R\omega/c)\cos(2R\omega/c) - \tfrac{1}{2}(R\omega/c)^2\sin(2R\omega/c)]$$

$$+ i[\tfrac{1}{2} - \tfrac{1}{2}\cos(2R\omega/c) - (R\omega/c)\sin(2R\omega/c) + \tfrac{1}{2}(R\omega/c)^2 + \tfrac{1}{2}(R\omega/c)^2 \cos(2R\omega/c)]\}. \tag{B1}$$

When $\omega \to \infty$ in the upper-half-plane, the dominant terms inside the curly brackets of Eq.(B1) will be $\pm\tfrac{1}{4}i(R\omega/c)^2 e^{-i2R\omega/c}$ (with $\pm$ for the cosine and sine terms, respectively). Consequently,

$$\Gamma(\omega) \cong i\varepsilon \left(\frac{Z_0 q^2}{24\pi}\right) e^{-i2R\omega/c}. \tag{B2}$$

The upper-half-plane poles of Eq.(17) (with $\beta = 0$) may thus be approximated as follows:

$$I\omega + \Gamma(\omega) \cong \tfrac{2}{3} m R^2 \omega + i\varepsilon \left(\frac{Z_0 q^2}{24\pi}\right) e^{-i2R\omega/c} = 0$$

$$\to \quad (i2R\omega/c)e^{i2R\omega/c} = \frac{Z_0 q^2 \varepsilon}{8\pi c m R} \quad \to \quad \omega_k = -\left(\frac{ic}{2R}\right) W_k\left(\frac{Z_0 q^2 \varepsilon}{8\pi c m R}\right). \tag{B3}$$

Here, $W_k(z)$ is the $k^{\text{th}}$ branch of the Lambert function,[10] defined over the complex $z$-plane as the inverse of the function $z = we^w$. The integer $k$ may assume positive, zero, and negative values. Although Eq.(B3) is only asymptotically valid in the limit when $\varepsilon \to 0$, numerical evaluations indicate its accuracy over a broad range of the parameters. For $k \neq 0$, the Lambert function has a singularity at the origin, its value approaching $-\infty$ as its argument goes to zero, which shows that the imaginary part of the upper-half-plane poles approaches $+\infty$ when $\varepsilon \to 0$. Thus, any departure from the full radiation reaction function $\Gamma(\omega)$ by way of attenuating its out-of-phase component will result in the dipole's transfer function $\Omega_0/T_0$ acquiring poles in the upper-half of the $\omega$-plane.

Equation (B3) also indicates that an increase in the inertial mass $m$ of the particle (for example, switching from $m_{\text{electron}}$ to $m_{\text{neutron}}$) reduces the argument of $W_k(\cdot)$, thus causing the poles in the upper-half of the $\omega$-plane to move further up, a behavior that is confirmed by numerical calculations.